\documentclass[12pt,fleqn]{article}
\usepackage{amsmath,amssymb,amsthm,cite}

\theoremstyle{definition}
\newtheorem{remark}{Remark}

\begin{document}
\allowdisplaybreaks

\begin{flushleft}
\Large \bf Symmetry, Equivalence and Integrable Classes\\
of Abel Equations
\end{flushleft}

\begin{flushleft} \bf
Vyacheslav M. BOYKO
\end{flushleft}

\noindent
{\small Institute of Mathematics of NAS of Ukraine, 3
Tereshchenkivska Str., Kyiv 4, 01601 Ukraine\\
E-mail: boyko@imath.kiev.ua}

\begin{abstract}
\noindent
We suggest an approach for description of integrable cases
of the Abel equations. It is based on increasing of the order of equations up to the second one
and using equivalence transformations for the corresponding second-order ordinary differential equations.
The problem of linearizability of the equations under consideration is discussed.
\end{abstract}

\section{Introduction}

A diversity of methods were developed to date for finding solutions
of nonlinear ordinary differential equations (ODE).
Everybody who encounters
integration of a particular ODE uses, as a rule, the
accumulated databases (or reference books) of the classes
of ODE and methods for integration of them (e.g.
\cite{Kamke1959,Polyanin&Zaitsev1995}).
But if an ODE does not belong to any of the described classes
then it does not mean that there is no
approaches for finding solutions of this ODE in a closed form.

The symmetry approach is one of the most algorithmic approaches for
integration and lowe\-ring of the order of ODE that admit a certain
nontrivial symmetry (see e.g.\ Lie's book \cite{Lie1891},
the books \cite{Ibragimov:1992,
Olver1986,Olver1995}
and review papers \cite{Ibragimov:1992,Schwarz2000}).
In the framework of the symmetry approach (and its modifications)
it is possible to obtain many of the known classes of integrable ODE.
However, the needs of the applications stimulate new research into
development of new methods for construction of ODE solutions
in the closed form. The papers
\cite{Abraham-Shrauner1996,Abraham-Shrauner2002,
Abraham-Shrauner&Guo1992,
Adam&Mahomed1998,Adam&Mahomed&Kara1994,
Berkovich2002,Berth&Czichowski2001,Bocharov&Sokolov&Svinolupov1993,
Cheb-Terrab&Roche2000,Cheb-Terrab&Roche2003,
Edelstein&Govinder&Mahomed2001,
Euler&Euler2004,
Euler&Wolf&Leach&Euler2003,
Gonzalez-Gascon&Gonzalez-Lopez1983,Gonzalez-Gascon&Gonzalez-Lopez1988,
Ibragimov:1992,Mahomed&Leach1989,
Mahomed&Leach1990,Mallin&Mahomed&Leach1994,
Olver1995,Polyanin&Zaitsev1995,
Popovych&Boyko2002,Popovych&Boyko&Nesterenko&Lutfullin2003,
Revenko&Boyko1995,
Schwarz1998,Schwarz1998a,
Schwarz2000} may give an idea of current developments and directions
of research in the field of symmetry (algebraic) methods for investigation of ODE.

The problem of finding Lie symmetries for the
first-order ODE is equivalent to finding solutions
for these equations, and for this reason the direct
application of the Lie method is complicated in the general case.
On of the well-known approaches in the cases when for
a given ODE it is not feasible (or not effective)
to apply the Lie method directly, is increasing of
the order of the ODE under consideration (in particular,
to obtain a second-order ODE related to the respective ODE
by a change of variables). For examples of utilisation
of such approach we can refer to papers
\cite{Abraham-Shrauner1996,
Abraham-Shrauner2002,
Abraham-Shrauner&Guo1992,
Adam&Mahomed1998,
Adam&Mahomed&Kara1994,
Edelstein&Govinder&Mahomed2001,
Mahomed&Leach1989,
Mahomed&Leach1990,
Mallin&Mahomed&Leach1994}. In such cases, if
the ``induced'' equation of a higher order admits
a non-trivial Lie symmetry (that generated a non-local symmetry
for the initial equation), we can speak
of so-called hidden symmetries for
an initial equation (for more details see \cite{Abraham-Shrauner1996,
Abraham-Shrauner2002,
Abraham-Shrauner&Guo1992}).

\section{Main results}

In this paper we study Abel equations having the form
\begin{gather}
\label{eq:01}
\dot p(f_5(y)p+f_0(y))=p^3f_4(y)+p^2f_3(y)+pf_2(y)+f_1(y),
\end{gather}
where $p=p(y)$, $\dot p=\frac{dp}{dy}$,
$f_i$, $i=0,\ldots,5$, are arbitrary smooth functions (with $f_1$, $f_2$, $f_3$, $f_4$, $f_5$
not identically vanishing simultaneously).
In view of existence of the
gauge transformation of multiplication
by an  arbitrary function of~$y$,
any equation \eqref{eq:01}
 can be reduced to one of the following canonical forms
(respectively, Abel equations of the first
and the second kind, see e.g. \cite{Abel1839,Kamke1959,Polyanin&Zaitsev1995}):
\begin{gather}
\dot p=p^3f_4(y)+p^2f_3(y)+pf_2(y)+f_1(y),\label{eq:0}
\\
\label{eq:1}
\dot p(p+f_0(y))=p^3f_4(y)+p^2f_3(y)+pf_2(y)+f_1(y).
\end{gather}
 Equations
\eqref{eq:0}, \eqref{eq:1} along with the Riccati equation are
among the ``simplest'' nonlinear first-order ODE that have extensive
applications. At the same time the problem of description of integrable
classes of these equations stays within the focus of current research, and
was previously considered in many papers
(see e.g.\ \cite{Adam&Mahomed1998,Adam&Mahomed&Kara1994,Cheb-Terrab2004,
Cheb-Terrab&Kolokolnikov2000,
Cheb-Terrab&Roche2000,Cheb-Terrab&Roche2003,
Mallin&Mahomed&Leach1994,
Polyanin&Zaitsev1995,
Popovych&Boyko2002,
Revenko&Boyko1995,
Schwarz1998,Schwarz1998a}).

Note that the Abel equations of the first and the second
kind \eqref{eq:0}, \eqref{eq:1} are related
with each other by a local change of variables
(namely, the equation \eqref{eq:1}
can be reduced to the form \eqref{eq:0}
by means of the change of variables $p=1/v(y)-f_0$).
Besides, the well-known Riccati equation is a partial
case of equation \eqref{eq:0}.

Further we will consider the following second-order ODE
\begin{gather}\label{eq:2-0}
\ddot y=\dot y^4f_4(y)+\dot y^3f_3(y)+\dot y^2 f_2(y)+\dot y f_1(y),\\
\label{eq:2}
\ddot y(\dot y+f_0(y))=\dot y^4f_4(y)+\dot y^3f_3(y)+\dot y^2 f_2(y)+\dot y f_1(y),
\end{gather}
where $y=y(x)$, $\dot y=\frac{dy}{dx}$, $\ddot y=\frac{d^2 y}{d x^2}$,
related to the Abel equations \eqref{eq:0} and \eqref{eq:1}.

The substitution $\dot y=p(y)$ reduces
equations \eqref{eq:2-0} and \eqref{eq:2}
respectively to the Abel equations~\eqref{eq:0}
and   \eqref{eq:1} (reduction of the order
for equations~\eqref{eq:2-0} and \eqref{eq:2}).
Such reduction is induced by the Lie
operator $X_1=\partial_x$ (that corresponds to invariance of equations
 \eqref{eq:2-0} and~\eqref{eq:2}
with respect to translations by the variable~$x$).
 This is exactly the fact that explains why
we consider equations~\eqref{eq:2-0} and  \eqref{eq:2}.

In the case when  \eqref{eq:2-0}
or \eqref{eq:2} are invariant with respect
to another operator (that is when~\eqref{eq:2-0}
or \eqref{eq:2} admit two-dimensional Lie algebras),
then equations \eqref{eq:2-0} and \eqref{eq:2}
are integrable in the framework of the Lie approach. And in
this way we can obtain exact solutions of the equations~\eqref{eq:0}
and \eqref{eq:1} respectively.

Further we will consider only the equation \eqref{eq:2}
(since equations \eqref{eq:0}--\eqref{eq:2} are interconnected~-- see Remark~3).
Let  \eqref{eq:2} admit a two-dimensional Lie algebra
\begin{gather}\label{eq:3}
L=\langle X_1,X_2\rangle, \qquad
X_1=\partial_x,\qquad X_2=\xi(x,y)\partial_x+\eta(x,y)\partial_y.
\end{gather}

We will consider a problem of description of inequivalent
equations \eqref{eq:2}
that are invariant with respect to
two-dimensional Lie algebras of the form  \eqref{eq:3}
(non-equivalent realizations of the operator $X_2$
in the algebra \eqref{eq:3} will determine canonical
representatives for equation \eqref{eq:2}).

It is well-known that any two-dimensional Lie algebra in the general
case, by means of choosing the basis operators $X_1$
and $X_2$ in an appropriate manner, may be reduced to
four nonequivalent cases~(see e.g.\ \cite{Ibragimov:1992,Ibragimov:1999,
Lie1891,Olver1986}).
In the framework of our problem additional
cases arise as we have fixed the form of the operator $X_1$.

So, it is quite straightforward to show that equation \eqref{eq:2} may admit
a two-dimensional Lie algebra~\eqref{eq:3} only of one of the following types:
\begin{gather}
1. \ \ [X_1,X_2]=0, \quad {\rm rank}\, L=1;\nonumber\\
2. \ \ [X_1,X_2]=0, \quad {\rm rank}\, L=2;\nonumber\\
3. \ \ [X_1,X_2]=X_1, \quad {\rm rank}\, L=1;\nonumber\\
4. \ \ [X_1,X_2]=X_1, \quad {\rm rank}\, L=2;\nonumber\\
5. \ \ [X_1,X_2]=X_2, \quad {\rm rank}\, L=1;\nonumber\\
6. \ \ [X_1,X_2]=X_2, \quad {\rm rank}\, L=2.
\label{eq:4}
\end{gather}

Further, utilising classification of two-dimensional algebras \eqref{eq:4},
we obtain that equation~\eqref{eq:2} may admit
only the following realizations of two-dimensional Lie algebras \eqref{eq:3}:
\begin{gather}
1. \ \ X_1=\partial_x,\quad X_2=\xi(y)\partial_x,\quad \xi(y)\not\equiv {\rm const};
\nonumber\\
2. \ \ X_1=\partial_x,\quad X_2=\xi(y)\partial_x+\eta(y)\partial_y,\quad
 \xi(y)\not\equiv {\rm const}
\ \mbox{or} \ \xi(y)\equiv 0, \quad \eta(y)\not=0;
\nonumber\\
3. \ \ X_1=\partial_x,\quad X_2=(x+\xi(y))\partial_x,\quad \xi(y)\not\equiv {\rm const}
\ \mbox{or} \ \xi(y)\equiv 0;
\nonumber\\
4. \ \ X_1=\partial_x,\quad X_2=(x+\xi(y))\partial_x+\eta(y)\partial_y,\quad
 \xi(y)\not\equiv {\rm const}
\ \mbox{or} \ \xi(y)\equiv 0, \quad \eta(y)\not=0;
\nonumber\\
5. \ \ X_1=\partial_x,\quad X_2=e^{x}\xi(y)\partial_x,\quad \xi(y)\not\equiv 0;
\nonumber\\
6. \ \ X_1=\partial_x,\quad X_2=e^{x}(\xi(y)\partial_x+\eta(y)\partial_y),
 \quad \eta(y)\not=0.\label{eq:5}
\end{gather}

It is clear that using these realizations we
can describe equations of the form \eqref{eq:2}
that are invariant with respect to two-dimensional Lie algebras
(similarly as we have discussed in \cite{Revenko&Boyko1995}).
However, this way is too cumbersome, and thus obtained types
of equations~\eqref{eq:2} will be quite complicated
(functions $f_i$, $i=0,\ldots,4$ in \eqref{eq:2} will
be expressed through coefficients of the
opera\-tor~$X_2$ from realizations \eqref{eq:5}).

It is straightforward to show that
the most general transformations that preserve the form
of the operator~$X_1$ we look as follows:
\begin{gather}\label{eq:6}
t=x+\omega(y),\qquad u=g(y),
\end{gather}
where $\omega(y)$, $g(y)$ are arbitrary smooth functions, $g(y)\not\equiv {\rm const}$.

After substitution \eqref{eq:6} equation \eqref{eq:2} takes the form
\begin{gather}
\ddot u((1-\omega'f_0)\dot u+f_0 g')g'{}^2=
\big(f_4-\omega''(1-\omega'f_0)-\omega'f_3+\omega'{}^2f_2-\omega'{}^3 f_1\big)\dot       u^4
\nonumber\\
\qquad{}+
\big(g'f_3-\omega'' g'f_0+g''(1-\omega'f_0)-2\omega' g' f_2+3\omega'{}^2g' f_1\big)\dot u^3
\nonumber\\
\qquad {}+ \big(g'{}^2f_2+g''g'{}^2f_0-3\omega' g'{}^2f_1\big)\dot u^2
+f_1g'{}^3 \dot u,\label{eq:2c}
\end{gather}
where $\omega'=\frac{d\omega}{dy}$, $\omega''=\frac{d^2 \omega}{dy^2}$,
$g'=\frac{dg}{dy}$, $g''=\frac{d^2 g}{dy^2}$
(in addition in \eqref{eq:2c} all functions
of the variable $y$ should be expressed as functions of the variable  $u$).

With $(1-\omega'f_0)\not\equiv 0$ equation \eqref{eq:2c}
belongs again to the class of equations \eqref{eq:2}.

\begin{remark}
With $(1-\omega'f_0)\equiv 0$ after the substitution \eqref{eq:6},
equation \eqref{eq:2} is transformed to the equation \eqref{eq:2-0},
that is reduced to the Abel equation of the first kind~\eqref{eq:0}.
\end{remark}

\begin{remark}
It is possible to regard that $(1-\omega'f_0)\not\equiv 0$
for the equation \eqref{eq:2} as a result of the substitution \eqref{eq:6}
(we attain that by combination of transformations \eqref{eq:6}).
\end{remark}

Thus \eqref{eq:6} are equivalence transformations for \eqref{eq:2},
and, besides, these transformations preserve the form of the operator
$X_1=\partial_x$ in the algebra \eqref{eq:3}.

\begin{remark}
So, the transformations \eqref{eq:6} are equivalence transformations
for the class of equations \eqref{eq:2-0}--\eqref{eq:2}.
Moreover, if we prolongate
these transformations for $\dot u=p$ then they
form an equivalence transformation group for \eqref{eq:01} and
include as a subgroup in the complete equivalence group
of class \eqref{eq:01}, which are formed by the transformations
\[
\tilde y= F(y), \quad \tilde p= \frac{P_1(t) p+Q_1(t)}{P_2(t) p+Q_2(t)},
\]
where
$F$, $P_1$, $P_2$, $Q_1$, $Q_2$ are arbitrary analytic functions, and $P_1Q_2-P_2Q_1\not=0$.
\end{remark}

Thus, by means of transformations \eqref{eq:6},
realizations \eqref{eq:5} of the algebra
\eqref{eq:3} may be reduced to the simplest canonical form.
The transformations \eqref{eq:6} in that process will not take us out of the class of
equations \eqref{eq:2}.

By means of transformations \eqref{eq:6}
the realizations \eqref{eq:5} of two-dimensional
Lie algebras \eqref{eq:3} admitted for equation \eqref{eq:2}
are reduced to the following canonical realizations:
\begin{gather}
1. \ \ X_1=\partial_x,\qquad X_2=y\partial_x;
\nonumber\\
2. \ \ X_1=\partial_x,\qquad X_2=\partial_y;
\nonumber\\
3. \ \ X_1=\partial_x,\qquad X_2=x\partial_x;
\nonumber\\
4. \ \ X_1=\partial_x,\qquad X_2=x\partial_x+y\partial_y;
\nonumber\\
5. \ \ X_1=\partial_x,\qquad X_2=e^{x}\partial_x;
\nonumber\\
6. \ \ X_1=\partial_x,\qquad X_2=e^{x}(\partial_x+\partial_y).\label{eq:5a}
\end{gather}

 In accordance to \eqref{eq:5a} we obtain
 the following  integrable
 cases for equation \eqref{eq:2} that
 are non-equivalent with respect to \eqref{eq:6}:
 \begin{gather}
 1. \ \ddot y=\alpha(y) \dot y^3;\nonumber\\
 2. \ \ddot y (\dot y+e)=d\dot y^4 +c\dot y^3+b\dot y^2+a\dot y;\nonumber\\
 3. \ \ddot y = \alpha(y)\dot y^2;\nonumber\\
 4. \ y \ddot y(\dot y+e)=d\dot y^4 +c\dot y^3+b\dot y^2+a\dot y;\nonumber\\
 5. \ \ddot y(\dot y+\beta(y))=\alpha(y)\dot y^3 +(1-\alpha(y)\beta(y))\dot y^2-
\beta(y)\dot y;\nonumber\\
 6. \ a) \ f_0=0:\nonumber\\
 \phantom{6. \ a)} \ \ddot y=d e^{y}\dot y^3+(-3 de^{y}+c)\dot y^2
 +(-de^{y}-(2c+1)+be^{- y})\dot y\nonumber\\
 \phantom{6. \ a)} \ \phantom{\ddot y=}{}+
  (-de^{y}+(c+1)-be^{- y}+a e^{-2 y});\nonumber\\
 \phantom{6.} \ b) \ f_0\not = 0:\nonumber\\
 \phantom{6. \ b)} \ \ddot y(\dot y+\alpha(y))=-\dot y^3+(1-\alpha(y))\dot y^2+
 \alpha(y)\dot y,\label{eq:2Cases}
 \end{gather}
where
$\alpha(y)$, $\beta(y)$ are arbitrary smooth functions, $a$, $b$, $c$, $d$, $e$ are constants.

The case 6a in \eqref{eq:2Cases} may be simplified by
means of the substitution $t=x$, $u=e^y$ (see
\eqref{eq:6} and \eqref{eq:2c}).

Equations \eqref{eq:2Cases} determine non-equivalent
cases of the form   \eqref{eq:2} that admit two-di\-men\-sio\-nal algebras
\eqref{eq:5a} up to equivalence transformations~\eqref{eq:6}.

Thus, summarising the above, we come to the following
scheme for integration of the Abel equation \eqref{eq:1}:
\begin{itemize}
\itemsep=0pt
\item we increase the order of equation \eqref{eq:1},
considering a second-order equation \eqref{eq:2};
\item if a corresponding equation \eqref{eq:2} admits
a two-dimensional Lie algebra, then we reduce this algebra
to one of the canonical forms
\eqref{eq:5a}, and thus the equation is reduced
to the respective canonical forms \eqref{eq:2Cases};
\item we integrate the canonical form \eqref{eq:2Cases};
\item making reverse changes of variables we obtain the
solution of the Abel equation \eqref{eq:1}.
\end{itemize}

\section{Case of Lie's linearization test}

According to results of S.~Lie \cite{lie1883} (see also \cite{Ibragimov:1992,Ibragimov:1999,
Ibragimov&Magri:2004, Ibragimov&Meleshko:2004})   second-order ODEs
\begin{gather}\label{eq3.1}
\ddot y=f(x,y,\dot y)
\end{gather}
can be reduced to the form
\begin{gather}\label{eq3.2}
\ddot u=0,
\end{gather}
by point change of variables
\begin{gather}\label{eq3.3}
t=\varphi(x,y),\quad u=\psi(x,y), \quad u=u(t)
\end{gather}
if  equations \eqref{eq3.1} is at most cubic in the first derivative, i.e. only
if equations \eqref{eq3.1} has the form
\begin{gather}\label{eq3.4}
\ddot y+F_3(x,y)\dot y^3+F_2(x,y)\dot y^2+F_1 (x,y)\dot y +F(x,y)=0,
\end{gather}
where
\begin{gather}
F_3(x,y)=\frac{\varphi_y\psi_{yy}-\psi_y\varphi_{yy}}
{\varphi_x\psi_y-\varphi_y\psi_x},\nonumber\\
F_2(x,y)=\frac{\varphi_x\psi_{yy}-\psi_x\varphi_{yy}+2(\varphi_y\psi_{xy}-\psi_y\varphi_{xy})}
{\varphi_x\psi_y-\varphi_y\psi_x},\nonumber\\
F_1(x,y)=\frac{\varphi_y\psi_{xx}-\psi_y\varphi_{xx}+2(\varphi_x\psi_{xy}-\psi_x\varphi_{xy})}
{\varphi_x\psi_y-\varphi_y\psi_x},\nonumber\\
F(x,y)=\frac{\varphi_x\psi_{xx}-\psi_x\varphi_{xx}}
{\varphi_x\psi_y-\varphi_y\psi_x}.\label{eq3.5}
\end{gather}
For given function $F_3(x,y)$, $F_3(x,y)$, $F_1(x,y)$ and $F(x,y)$  linearization is possible
iff the over-determined system \eqref{eq3.5} is integrable.
S. Lie proved that system \eqref{eq3.5} is integrable iff the following auxiliary system
for $w$ and $z$
\begin{gather}
\frac{\partial w}{\partial x}=zw-FF_3-\frac 13 \frac{\partial F_1}{\partial y}+
\frac 23 \frac{\partial F_2}{\partial x},\nonumber\\
\frac{\partial w}{\partial y}=-w^2+F_2w+F_3 z+ \frac{\partial F_3}{\partial x}-F_1F_3,\nonumber\\
\frac{\partial z}{\partial x}=z^2-Fw-F_1 z+ \frac{\partial F}{\partial y}+FF_2,\nonumber\\
\frac{\partial z}{\partial y}=-zw+FF_3-\frac 13 \frac{\partial F_2}{\partial x}+
\frac 23 \frac{\partial F_1}{\partial y}
\label{eq3.6}
\end{gather}
is compatible. The compatibility conditions for this system have the form
\begin{gather}
3(F_3)_{xx}-2(F_2)_{xy}+(F_1)_{yy}=(3F_1F_3-F_2^2)_x-3(FF_3)_y- 3F_3F_y+F_2(F_1)_y,\nonumber\\
3F_{yy}-2(F_1)_{xy}+(F_2)_{xx}=3(FF_3)_x-3(FF_2-F_1^2)_y+ 3F(F_3)_x-F_1(F_2)_x
\label{eq3.7}
\end{gather}
(subscripts $x$ and $y$ denote differentiations with respect to $x$ and $y$, respectively).

So, following \cite{Ibragimov&Magri:2004,Ibragimov&Meleshko:2004}
a necessary and sufficient condition
of lineariziation for equations of form \eqref{eq3.4} is that
functions $F_3(x,y)$, $F_2(x,y)$, $F_1(x,y)$ and $F(x,y)$ satisfy the conditions \eqref{eq3.7}.

In case $f_0(y)\equiv 0$ equations \eqref{eq:2} is partial case of \eqref{eq3.4}, i.e.
have the following form
\begin{gather}
\label{eq3.8}
\ddot y=\dot y^3f_4(y)+\dot y^2f_3(y)+\dot y f_2(y)+ f_1(y).
\end{gather}
This equations can be linearizable if
$f_4(y)$, $f_3(y)$, $f_1(y)$ and $f_1(y)$ satisfy the conditions (following \eqref{eq3.7})
\begin{gather}
(f_2)_{yy}=3(f_1f_4)_y+ 3f_4(f_1)_y-f_3(f_2)_y,\nonumber\\
3(f_1)_{yy}=3(f_1f_3-f_2^2)_y.
\label{eq3.9}
\end{gather}
And point transformations \eqref{eq3.3} which linearizing \eqref{eq3.8} can
be found from system \eqref{eq3.5}.

Let us note that a second-order ODE is linearizable iff it admits
an eight-dimensional Lie algebra.
So, any linearizable equation \eqref{eq3.8} belongs, up to equivalence
 transformations \eqref{eq:6},
belong to the set of equations \eqref{eq:2Cases}.

\section{Conclusion}

It is obvious from the above that there is an alternative
way for generation of new integrable cases of the Abel
equation based on utilisation of the relation
between the Abel equations of the first and the second kind,
and relation between the equations \eqref{eq:2-0}
and \eqref{eq:2} by means of the transformations \eqref{eq:6}.
Thus, starting from some integrable Abel equation (that is of such
equation for which the solution is known) it is possible
to obtain new integrable cases of the Abel equations (solutions
of these equations will be
related through transformations~\eqref{eq:6}).
It would be possible to use for this purpose even the
well-known Riccati equation that is a partial case of the equation~\eqref{eq:0}
(for generation of integrable Riccati equations an approach
that is proposed in \cite{Popovych&Boyko2002} may be used).

We hope that new
results for classification of integrable classes of ODE may be obtained  also using our
classification of inequivalent realizations
of real low-dimensional Lie algebras~\cite{Popovych&Boyko&Nesterenko&Lutfullin2003}.

\subsection*{Acknowledgements}

The author are grateful to
Roman Popovych and Irina Yehorchenko
for useful discussions and interesting comments.
the research was partially supported by National Academy of Science of Ukraine
in the form of the grant for young scientists.

\end{document}